# Experiment and Modeling of Rice Winnowing: Granular Segregation Method in Ancient Traditions

Rahmawati Munir[1], Handika Dany Rahmayanti[1], Riri Murniati[1], Dui Yanto Rahman[1], SparisomaViridi[1] and Mikrajuddin Abdullah[1,a]

[1]*Department of Physics, Bandung Institute of Technology*
*Jalan Ganesa 10 Bandung 40132, Indonesia*
[(a)]*Email: mikrajuddin@gmail.com*

**ABSTRACT**

Rice winnowing is a process of separation of small and large rice grains by air flow practiced since the ancient human history especially in societies where rice is the main source of carbohydrate (in Asia, Africa, and Latin America). Indeed, this process contains rich of scientific rule but has never been documented by the old society. We report here experimental investigation of the rice winnowing and develop a physical model to explain the process of segregation of rice grains having different size or density. Flapping the tray in the winnowing process, generates a vortex centered at position around the tray free end. We demonstrated numerically that the effectiveness of segregation is strongly depended on the different in grain sizes (for grain from the same material), the initial position of the grain, and the angular velocity of the vortex generated by flapping the tray. We obtained a phase diagram describing different final conditions of winnowing process (either the small grains move towards the tray fee end or move toward the inner end of the tray, able or unable to leave the tray at the free end). The result can be useful to design a new method in separating grains based on size or density.

**INTRODUCTION**

Rice winnowing is a process of separation of small and large rice grains by air flow. Different accelerations experienced by the grains result in different locations where the grains touch the surface, depending on size or density. The grains of different sizes are then collected from the surface after well separated.

In open spaces, such as in paddy fields, winnowing process takes advantages of natural wind blowing at the place. This process is often used by farmers who are harvesting rice to separate the filled grains from empty grains [1,2] or unwanted matters like weeds, straws, sand,dust particles etc. to give cleaner outcome [3]. The empty grains, for example, will be thrown far away due to large acceleration (large force accompanied by small mass). But it is often the natural wind velocity may beunfavorable for the winnowing process thereby increasing operating time and difficulties in separating the unwanted matters [4]. To the contrary, in close spaces where no natural wind blows, the worker must generate air flow by him/herself. And the ancient common method for producing air flow is by flapping a tray above which the rice grains are placed. We can watch several videos regarding rice winnowing [5-14]

It is very clear that the rice winnowing is an example of traditional methods, and has been practiced thousands of years, especially in societies where rice is the main source of carbohydrate (in Asia, Africa, and Latin America). Indeed, the mechanism of rice winnowing sounds scientifically, although it has never been documented by the old communities. It contains physical processes, starting from generating vortex by flapping the tray. The vortex produces different acceleration to grains of different sizes or densities so that different grains move at different path. Grains of different sizes are slightly separated on the tray surface. By repeating the tray flapping, the grains with different sizes segregate progressively, and finally they can be collected separately.

The discussion of motion of particles in two-dimensional vortex has been reported by Lecuona et al [15]. They measured the trajectory of small particle (size of around 1 μm) and the density ratio between fluid and particle is of the order of $10^{-3}$. This density ratio is comparable to that of air and rice grain. However, to best of our knowledge, development of physical mechanism underlying the process of rice winnowing (the grain sizes are in millimeters) by flapping tray were nearly unpublished.

The purpose of this paper is to develop a simple physical foundation for describing the grain segregation in the process of rice winnowing by tray flapping. Simple experimental was also conducted to enrich the explanation. Of course, this topic is not a breakthrough physical problem, instead it puts a scientific explanation to an ancient (tradition) method that has been practiced in different continents for thousands of years ago. This is might be considered as ethnoscience or ethnophysics. As explained by Vlaardingerbroek, ethnoscience is the study of knowledge in its cultural context as a cultural adaptation to the world in which the people practicing in live [16]. Simulation of rice winnowing in a machine has been reported by Shrestha et al [17], but they focused on open space winnowing and apply simple air friction to rice grains. Both numerical and analytical investigations will be discussed.

**EXPERIMENTAL**

**Experimental Procedures**

We first demonstrate the effect of flapping on the segregation of mixtures of grains from two different sizes. The grains composed of sorted rice grains of 1.65 mm and 4.05 mm in diameters, appeared in different colors. The grains were placed on a bamboo tray with a diameter of around 0.6 m. Initially the grains are mixed homogeneously on the tray surface. The image of the grain after touching the tray surface for each flapping was recorded. We also used a mixture of soybean (large size) + mung bean (small size) and basil seeds (small size) + grain for making cake (large size) to show the segregation also happens on grains from different materials.

We also observed the flow of air above the tray when flapping the tray. For this purpose, we made a hole at the tray bottom that fitted with a small and flexible plastic hose. One end of the hose is glued at the tray bottom and the other end was connected to a smoke generator. The smoke was then transported through the hose and exit on the tray bottom. When flapping the tray, the flow pattern of the smoke was recorded.

**Experimental Results and Discussion**

Figure 1 shows the spatial distribution of a mixture of small (red color) and large (green color) rice grains on the tray surface after several times of flapping: (a) once, (b) twice, (c) 3 times, (d) 4 times, (e) 5 times, (f) 6 times, (g) 7 times, (h) 8 times (i) 9 times and (j) 10 times. It is clear that the small grains progressively move toward the front edge (free edge) of the tray when increasing the number of flapping. It is indicated by displacement of red color to the tray front edge. The green color spatial distribution (representing large grains), on the other hand, seems to be unchanged.

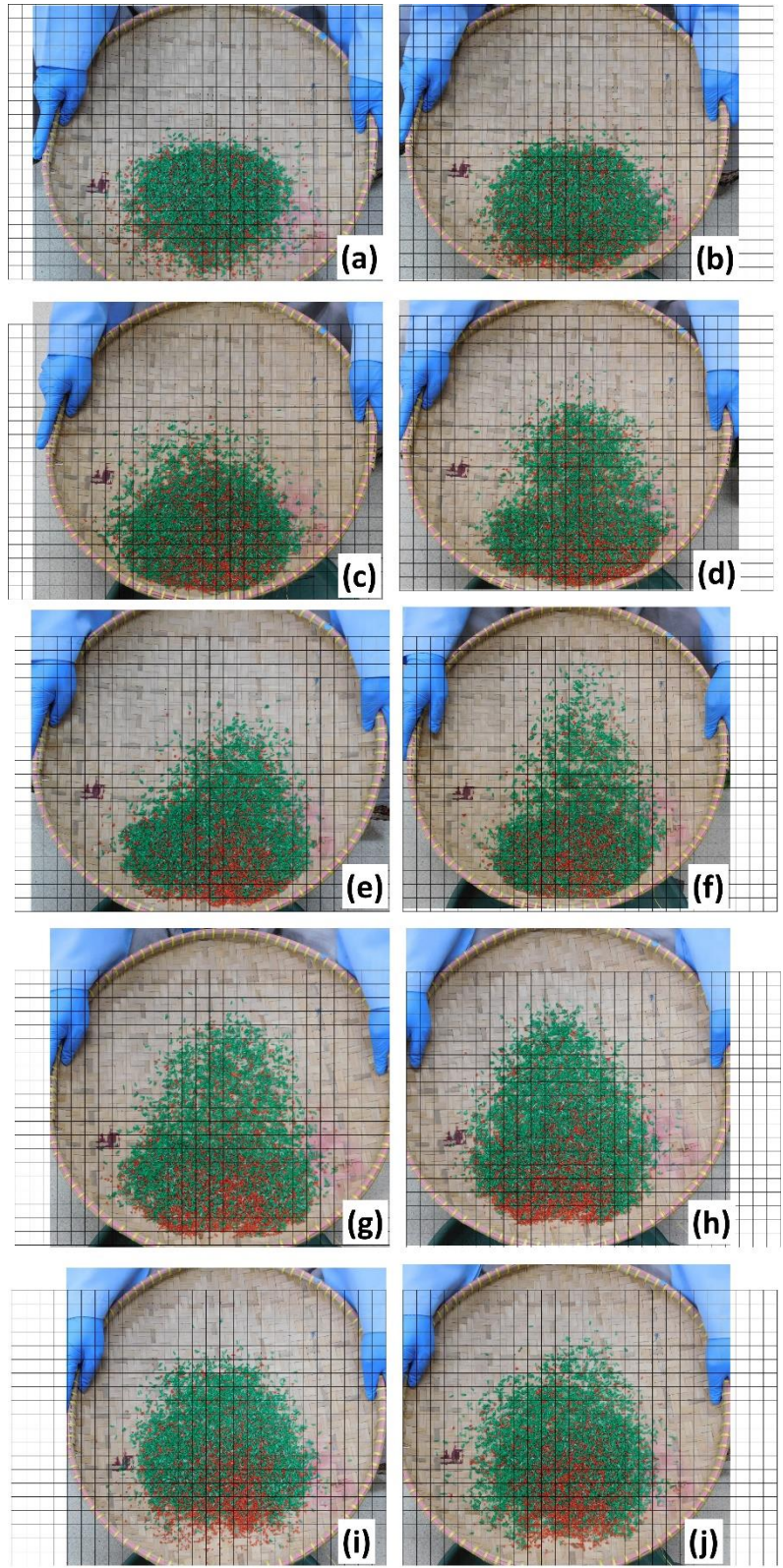

**Figure 1** Spatial distribution of mixtures of small (red color) and large (green color) rice grains on the tray surface after several times of flapping: (a) once, (b) twice, (c) 3 times, (d) 4 times, (e) 5 times, (f) 6 times, (g) 7 times, (h) 8 times (i) 9 times and (j) 10 times

Figure 2 shows the distribution of mixtures of two-sized grains before (left) and after (right) flapping the tray several times. (a1) and (a2) are mixtures of soybean grains (large, brown) and mung bean grains (small, green), while (b1) and (b2) are mixtures of basil seeds (small, black) and grains for making cake (large, white). It is also very clear that although initially two different grains were mixed homogeneously, they then separate after flapping the tray several times.

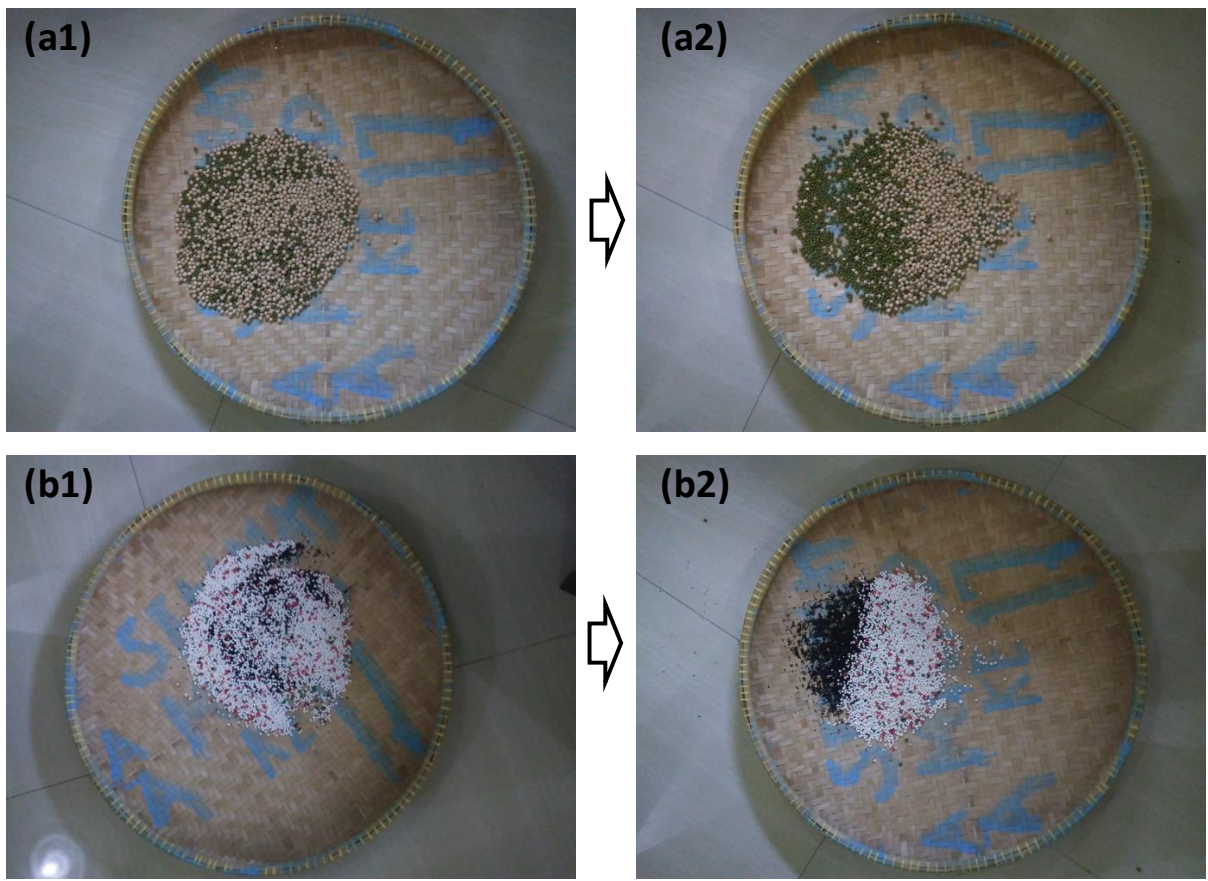

**Figure 2** Spatial distribution of mixtures of two-sized grains from different materials before flapping (left) and (right) after flapping for several times. (a1) and (a2) are mixtures of soybean grains (large, brown) and mungbean grains (small, green), while (b1) and (b2) are mixtures of basil seeds (small, black) and a grain for making cake (large, white).

We also counted the number distribution of small and large rice grains on the tray surface after once, 6 times, and 8 times of flapping. We divide the tray surface into identical squares as displayed in Fig. 3(a). We then counted the number of either small or larger grains located at a certain distance from the tray front edge. Figure 3(b) shows the counting of small rice grains against position at different flapping times and Fig.3(c) is the corresponding figure for large grains. We can clearly see that the position of small grains progressive move toward the tray front edge, while the position of large grains relatively unchanged up to 8 times flapping. Therefore, the segregation clearly developed when increasing the times of flapping.

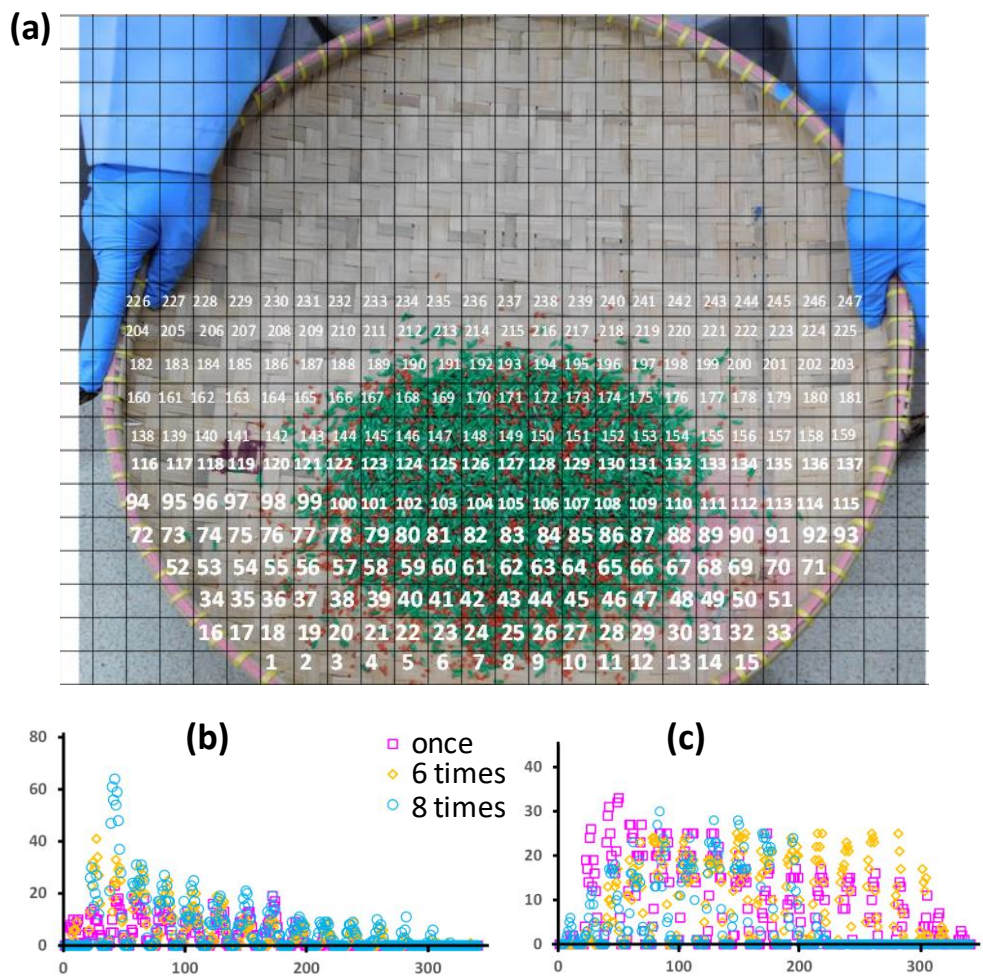

**Figure 3** (a) Dividing the tray surface into identical squares to calculate the grain spatial distribution, (b) counting the distribution of small grains, and (c) counting the distribution of small grains, (square) is the distribution after once flapping, (diamond) is the distribution after 6 times flapping and (circle) is the distribution after 8 times flapping. The zero coordinate for horizontal axis means the front edge of the tray.

Figure 4 shows the trajectories of smoke (circles), rice grain of 1.65 mm in diameter (square) and rice grain of 4.05 mm in diameter (triangle) when flapping the tray. We observed the smoke (air) circulate counter clockwise near the end of the tray. Dashed circle is the fitting for symbols of the air trajectory to confirm the circular trajectory of the air. Therefore, flapping of the tray creates a vortex around the front edge of the tray. The vortex is generated above the tray surface when flapping down the tray. Indeed, there are many simulations showing the development of such a vortex when describing the flapping wing [18-22]. Flapping the tray can be considered as an example of flapping wing process. This vortex is responsible for separating small and large grains. It is also clear that the grains are located to the left of the vortex center, to mean that when taking the vortex center as the frame origin and the $x$ axis direct rightward, the grains are located at the negative $x$ positions.

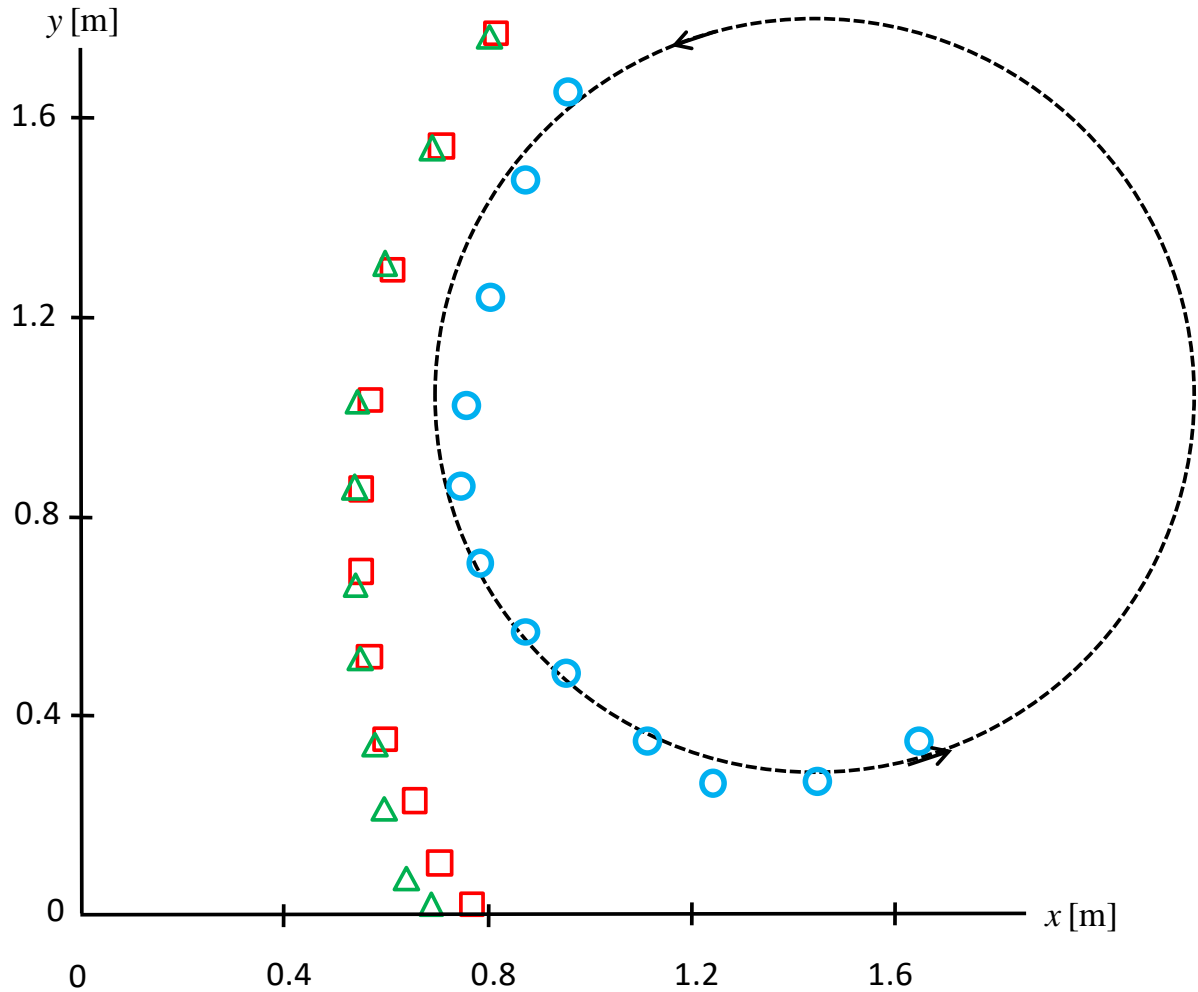

**Figure 4** Trajectories of air (circle), small rice grains (square), and large rice grain (triangle) after flapping down the tray. Dashed circle is the fitting for the air trajectory to confirm the shape is circular.

## MODELLING

### Model Development

For simplicity, we assume the grains are moving in a fluid that satisfies a dilute limit condition, where only coupling between particles and air exists. We will consider that the particle motion does not affect the fluid flow, although this is very drastic simplification for our case since the density of rice grain when flapping is high so that the air flow is affected. But in the present model we ignore it so that the results might slightly different from the measurement result. However, we assume that qualitatively the conclusion remains unchanged.

The equation for particle motion in dynamical fluid can be expressed in any fashions. At present, we will use the relationship introduced by Maxey and Riley [23,24] to describe the evolution of granular position in a dilute concentration as

$$V\rho_p \frac{d\vec{v}_p}{dt} = 3\pi\phi\mu\left[\vec{u}(\vec{x},t)\big|_{\vec{x}=\vec{x}_p(t)} - \vec{v}_p(t)\right] + V(\rho_p - \rho_a)\vec{g} + V\rho_a \frac{D\vec{u}}{Dt}\bigg|_{\vec{x}=\vec{x}_p(t)}$$

$$+ \frac{1}{2}V\rho_a \frac{d}{dt}\left[\vec{u}(\vec{x},t)\big|_{\vec{x}=\vec{x}_p(t)} - \vec{v}_p(t)\right] - \frac{3}{2}\pi\phi^2\mu\int_0^t \left\{\frac{\frac{d}{d\tau}\left[\vec{v}_p(\tau) - \vec{u}(\vec{x},t)\big|_{\vec{x}=\vec{x}_p(t)}\right]}{\sqrt{\pi\mu(t-\tau)\rho_a}}\right\} d\tau$$

(1)

with $\vec{v}_p$ the particle velocity, $\vec{u}$ the air velocity, $\rho_p$ the particle density, $\rho_a$ the air density, $\mu$ dynamic viscosity of air, $\phi$ the particle diameter, $V = \pi\phi^3/6$ the particle volume by assuming the particle has spherical shape, and $\vec{x} = x\hat{i} + y\hat{j}$ the instantaneous particle position (the motion of granules is two dimensional). The last term is the Bassel history term that has been proven to initially decay as $t^{-1/2}$ [15] and later $t^{-2}$ so that it might be neglected by assuming the time of process is much larger than the characteristic time for decay [25].

Equation (1) can be easily solved if we know the velocity of air. For this purpose, we recorded the motion of air only when the tray is flapped as if winnowing the rice. Similar to that illustrated in Fig. 4, the air circulates counter clockwise. By approximating the angular velocity of the air to be a constant, the air velocity components in the vortex core can be written as

$$u_x = -\Omega y \quad \text{and} \quad u_y = \Omega x \quad (2)$$

By carefully inspecting the space at which the air circulates we concluded that the radius of vortex is so large so that the rice grain can be considered to move in the vortex core. Therefore we ignore from consideration the motion of air outside the vortex core.

Now, let us simplify Eq. (1) by separating into components. After ignoring the Bassel history term, Eq. (1) can be written as

$$V\rho_p\left(1 + \frac{\rho_a}{2\rho_p}\right)\frac{d\vec{v}_p}{dt} = 3\pi\phi\mu\left[\vec{u}(\vec{x},t)\big|_{\vec{x}=\vec{x}_p(t)} - \vec{v}_p(t)\right] + V(\rho_p - \rho_a)\vec{g} + V\rho_a \frac{D\vec{u}}{Dt}\bigg|_{\vec{x}=\vec{x}_p(t)}$$

$$+\frac{1}{2}V\rho_a \frac{d}{dt}\vec{u}(\vec{x},t)\Big|_{\vec{x}=\vec{x}_p(t)} \tag{3}$$

It is easy to show that

$$\frac{D\vec{u}}{Dt} = -\Omega^2(x\hat{i} + y\hat{j}) \tag{4}$$

and

$$\frac{d\vec{u}}{dt} = -\Omega(\dot{y}\hat{i} - \dot{x}\hat{j}) \tag{5}$$

after assuming that the transient behavior dies quickly so that we take $\partial \vec{u}/\partial t = 0$. Furthermore, let us define

$$\varepsilon = \frac{\rho_a}{\rho_p} \tag{6}$$

Substitution Eqs. (4)-(6) into Eq. (3) and rearranging one obtains

$$\left(1+\frac{\varepsilon}{2}\right)\ddot{x} = -\frac{18\mu}{\rho_p \phi^2}\dot{x} - \frac{1}{2}\varepsilon\Omega\dot{y} - \varepsilon\Omega^2 x - \frac{18\mu\Omega}{\rho_p \phi^2}y \tag{7}$$

$$\left(1+\frac{\varepsilon}{2}\right)\ddot{y} = \frac{1}{2}\varepsilon\Omega\dot{x} - \frac{18\mu}{\rho_p \phi^2}\dot{y} + \frac{18\mu\Omega}{\rho_p \phi^2}x - \varepsilon\Omega^2 y - (1-\varepsilon)g \tag{8}$$

after assuming the grains have spherical shape.

Let us further simplify Eqs. (7) and (8) by substituting the data of the rice grain density. By characterizing 23 paddy varieties, Battacharya et al observed the average mass density of paddy grains of around 1,224 kg/m³ while the average density of rice grains of around $\rho_p \approx$ 1,452 kg/m³[26]. The mass density of fluid (air) $\rho_a \approx$ 1 kg/m³ so that $\varepsilon = 0.0007 \ll 1$, $1+\varepsilon/2 \approx 1$, and we can simplify Eqs. (7) and (8) as

$$\ddot{x} = -\frac{18\mu}{\rho_p \phi^2}\dot{x} - \frac{1}{2}\varepsilon\Omega\dot{y} - \varepsilon\Omega^2 x - \frac{18\mu\Omega}{\rho_p \phi^2}y \tag{9}$$

$$\ddot{y} = \frac{1}{2}\varepsilon\Omega\dot{x} - \frac{18\mu}{\rho_p \phi^2}\dot{y} + \frac{18\mu\Omega}{\rho_p \phi^2}x - \varepsilon\Omega^2 y - g \tag{10}$$

Equations (9) and (10) can be further simplified by introducing a dimensionless time variable

$$\tau = \Omega t \tag{11}$$

so that

$$\ddot{x} = -\frac{1}{St}\dot{x} - \frac{1}{2}\varepsilon\dot{y} - \varepsilon x - \frac{1}{St}y \tag{12}$$

$$\ddot{y} = \frac{1}{2}\varepsilon\dot{x} - \frac{1}{St}\dot{y} + \frac{1}{St}x - \varepsilon y - \frac{g}{\Omega^2} \tag{13}$$

where the derivation is with respect to $\tau$ and

$$St = \frac{\rho_p \Omega}{18\mu}\phi^2 \tag{14}$$

is the Stoke number. The angular velocity of vortex observed from the recorded video is around one circulation per second or $\Omega \approx 2\pi$ rad/s so that the estimated Stokes number is $St \approx$ 348 for common rice grains. This value is much larger than that investigated by Lecuona et al of about $10^{-3}$ [15].

The second order differential equations, Eqs. (12) and (13), can be solved analytically, and will be discussed at the end of this paper. At first, we will solve Eqs. (12) and (13) numerically. The numerical procedures are very easy by just using the simple numerical process. For this purpose, we slice the grain time of flight into a number of time steps. Suppose the time step is $\Delta\tau$. The derivation can be approximated as $\dot{x}_i = (x_{i+1} - x_i)/\Delta\tau$, $\ddot{x}_i = (x_{i+2} - 2x_{i+1} + x_i)/\Delta\tau^2$, $\dot{y}_i = (y_{i+1} - y_i)/\Delta\tau$, and $\ddot{y}_i = (y_{i+2} - 2y_{i+1} + y_i)/\Delta\tau^2$. The initial condition is $x_0$, $y_0$, $\vec{v}_0 = 0$. The zero initial velocity results $\dot{x}_0 = (x_1 - x_0)/\Delta\tau = 0$ to imply $x_1 = x_0$ and $\dot{y}_0 = (y_1 - y_0)/\Delta\tau = 0$ to imply $y_1 = y_0$. In discrete form Eqs. (12) and (13) become

$$x_{i+2} = 2x_{i+1} - x_i - \left[\frac{1}{St}(x_{i+1} - x_i) + \frac{1}{2}\varepsilon(y_{i+1} - y_i)\right]\Delta\tau - \left(\varepsilon x_i + \frac{1}{St}y_i\right)\Delta\tau^2 \tag{15}$$

$$y_{i+2} = 2y_{i+1} - y_i + \left[\frac{1}{2}\varepsilon(x_{i+1} - x_i) - \frac{1}{St}(y_{i+1} - y_i)\right]\Delta\tau + \left(\frac{1}{St}x_i - \varepsilon y_i - \frac{g}{\Omega^2}\right)\Delta\tau^2 \tag{16}$$

**Model Results and Discussion**

Equations (15) and (16) were used for simulations the traces of the grains. Based on observation of smoke circulation we estimated the angular velocity of $\Omega \approx 2\pi$ rad/s. Battacharya et al have measured the average grain length, width, and thickness of 23 paddy varieties and found 6.14-10.84 mm, 2.28 - 3.50, and 1.59 – 2.26 mm, respectively [26]. Fang et al investigated the *japonica* variety *kuanyejing* and found the average length is 6.93 mm and the average width is 3.16 mm [27]. Based on those findings, it is then reasonable if in this simulation we used the range of grain diameter between 2 mm to 8 mm.

Figure 5 shows the path of grains of different sizes: 2 mm, 4 mm, 6 mm, and 8 mm started at different initial positions, $x_0$ was fixed at 0.4 m while $y_0$ was varied : (A) 0.2 m, (B) 0.1 m, (C) 0, and (D) -0.1 m. Figures 5(A) and (B) demonstrated that when $y_0 > 0$, the smaller grains touched the tray surface at locations far away the tray front edge. The distance of the location where the grains touch the surface from the tray front edge increases. Therefore, the small grain will segregate to the inner part of the tray while the larger grain will segregate toward the tray front edge. This condition is opposite to the situation in the practical rice winnowing. The contrary situation was observed when $y_0 < 0$. Smaller grains will segregate toward the tray front edge. This is the condition usually observed in rice winnowing practice. Smaller grain will reach the front edge of the tray and some will leaved the tray front edge. The separation of smaller grains increases if $y_0$ is far below zero. At that initial position, the grains experience very large horizontal component of air velocity toward the tray front edge.

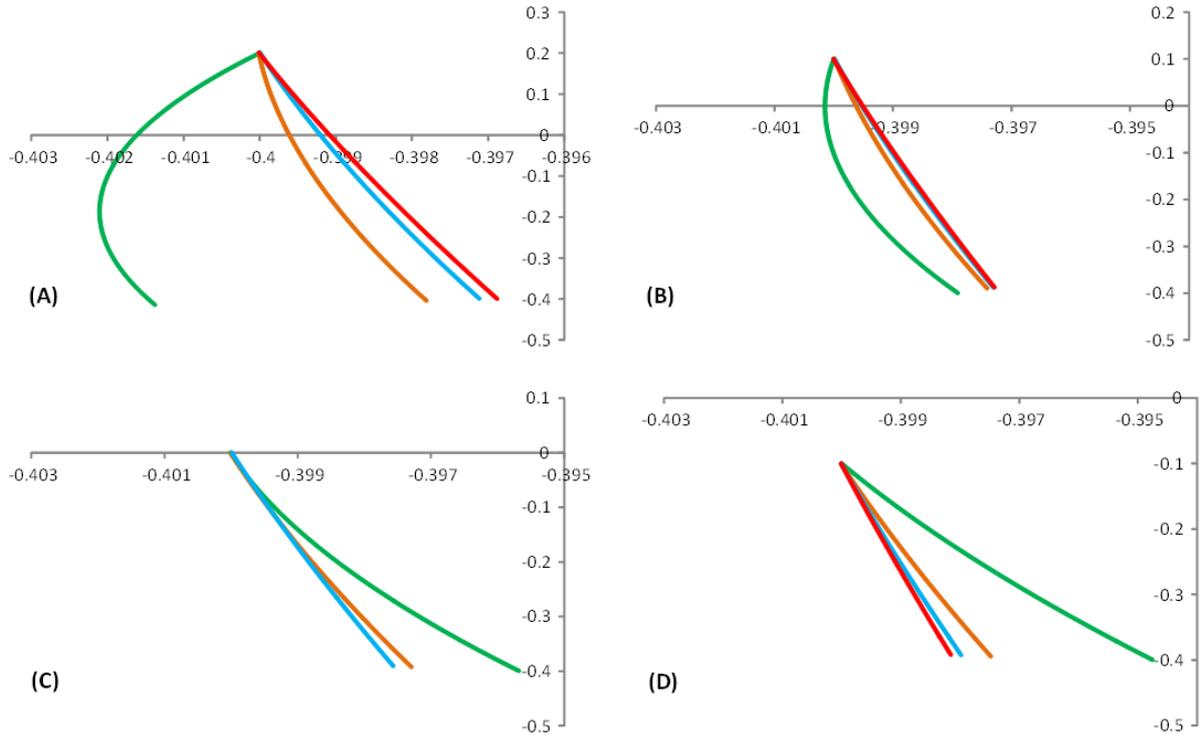

**Figure 5** The trajectories of grains of different sizes: (green) 2 mm, (brown) 4 mm, (blue) 6 mm, and (red) 8 mm started at different initial positions, $x_0$ was fixed at 0.4 m while $y_0$ was varied: (A) 0.2 m, (B) 0.1 m, (C) 0, and (D) -0.1 m. The angular velocity of the vortex was fixed at $\Omega = 4\pi$ rad/s.

The grain path is also depending on the $x_0$ as well as $y_0$. The dependence of $y_0$ has been demonstrated in Fig. 5. In Fig. 6, we show the trajectories of grains with 4 mm size started at different $x_0$: 0.2 m, 0 m, -0.2 m, and -0.4 m. We demonstrated for two $y_0$ values: 0.2 m and 0 m. We show that if $x_0$ increases, the displacement of grain toward the tray front edge decreases. Specifically, when $x_0$ is large positive, the grains eventually displaces toward the inner part of the tray. By increasing $y_0$, the trend of the grain to move to the inner part of the tray increases. From this observation we may concluded that to segregate the small grain to the front edge of the tray, we must reduce both $x_0$ and $y_0$. It is better to select both $x_0$ and $y_0$ are large negative.

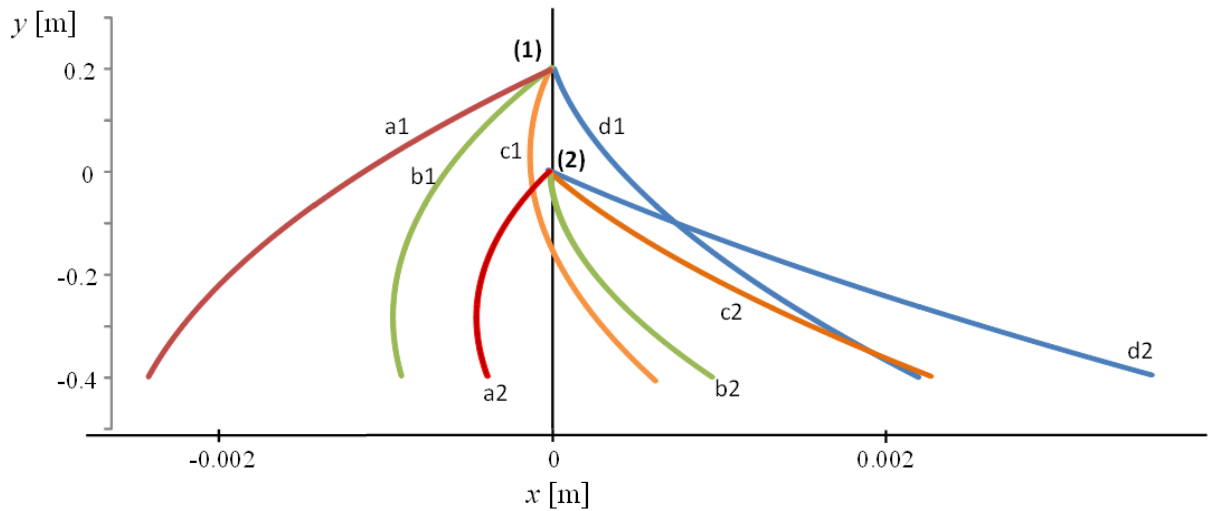

**Figure 6** The trajectories of grain of size 4 mm at different initial $x_0$ positions: (a1,a2) 0.2 m, (b1,b2) 0, (c1,c2) -0.2, and (d1,d2) -0.4 m and the $y_0$ coordinates were fixed at (1) 0.2 m and (2) 0. The angular velocity of the vortex was fixed at $\Omega = 4\pi$ rad/s.

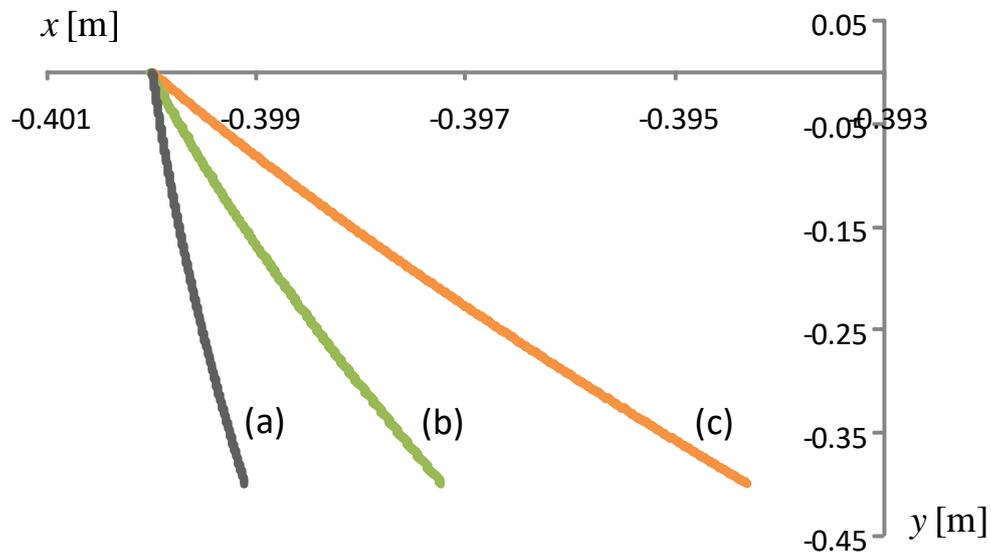

**Figure 7** The trajectories of grain of size 4 mm at different angular velocities of vortex $\Omega$: (a) $2\pi$ rad/s, (b) $4\pi$ rad/s, and (c) $6\pi$ rad/s. The initial position $x_0$= -0.4 m and $y_0 = 0$ were fixed.

We also explore the effect of vortex angular velocity of the grain trajectory. Fig. 7 shows the trajectories of grains of 4 mm size moving in a vortex core having different angular velocities: (a) $\Omega = 2\pi$ rad/s, (b) $\Omega = 4\pi$ rad/s, and (c) $\Omega = 6\pi$ rad/s. In simulations we have fixed the initial position at $x_0 = -0.4$ m and $y_0 = 0$. It is clear that the horizontal displacement of the grain increases as the angular velocity increases at this initial position. In practical the person flaps down quickly the tray to make a well segregation occur within a few times of flapping. The space swept by the tray becomes nearly empty when flapping down the tray. This space sonly filled by air from above to generate circular motion (vortex). When flapping down the tray quickly, the air will fill the empty space quickly so that the air move factor to generate large angular momentum.

Let us then inspect the paths of grains when the tray is flapped several times. This is the situation of normal rice winnowing by the farmer. One flapping is not enough to separate the grains. Figure 8 shows the traces of two grains of different sizes (2 mm and 4 mm) flapped for three times. Initially both grains were located at the same position ($x_0 = -0.4$ m and $y_0 = 0$ m). It is clearly observed that the displacement of small grains is larger than that of the larger grains. The first flapping produced very small displacement different between grains of different sizes. The different in displacement progressively increases when increasing the number of flapping. This is the reason why in the rice winnowing, the tray is flapped several times until different grain size can be separated manually.

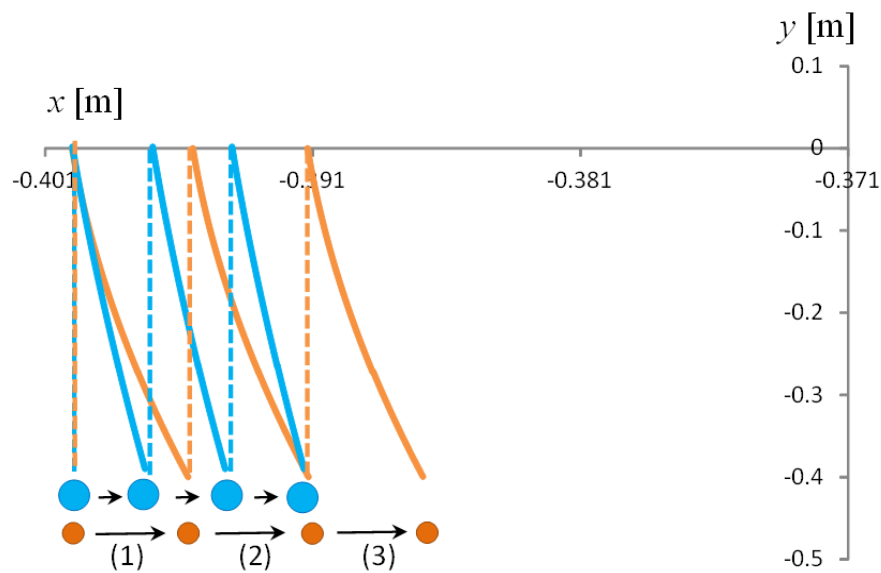

**Figure 8** The traces of two grains of different sizes, (brown) 2 mm and (blue) 4 mm, flapped for three times. Initially both grains were located at the same position ($x_0 = -0.4$ m and $y_0 = 0$ m). The angular velocity of the vortex was fixed at $\Omega = 4\pi$ rad/s.

## ANALYTICAL SOLUTION

To solve Eqs. (12) and (13) analytically, we can firstly transform them into one compact equation by introducing a complex variable

$$z = x + iy - \frac{ig}{\Omega^2(i/St - \varepsilon)} \tag{17}$$

Using this transformation, Eqs. (12) and (13) change into one single equation

$$\ddot{z} + \left(\frac{1}{St} - \frac{i}{2}\varepsilon\right)\dot{z} + \left(\varepsilon - \frac{i}{St}\right)z = 0 \tag{18}$$

Then by using a trial function $z = e^{\lambda t}$ we obtained a general solution of $z = A_1 e^{\lambda_1 t} + A_2 e^{\lambda_2 t}$, with

$$\lambda_{1,2} = \frac{-b \pm \sqrt{b^2 - 4c}}{2} \tag{19}$$

and

$$b = \left(\frac{1}{St} - i\frac{\varepsilon}{2}\right) \tag{20}$$

$$c = \left(\varepsilon - \frac{i}{St}\right) \tag{21}$$

Before proceeding the discussion, let us simplify some parameters by substituting the known quantities used in the experiment. The mass density of rice grains is around $\rho_P \approx 1{,}245$ kg/m$^3$, the mass density of fluid (air) $\rho_a \approx 1$ kg/m$^3$, the viscosity of air at room temperature $\mu \approx 2 \times 10^{-5}$ Pa s, the angular velocity of air (based on observation) is around $\Omega \approx 2\pi$ rad/s. The diameter of rice grain is around 4 mm. Based on this data we obtain the approximated values $\varepsilon \approx 0.0008$ and $St \approx 348$. Substituting those approximated values we can approximate other parameters as $b \approx 1/St$, $c \approx -i/St$, $b^2 - 4c \approx 1/St^2 + 4i/St \approx 4i/St$, $\sqrt{b^2 - 4c} \approx \sqrt{2/St}(1+i)$, $\lambda_1 \approx (-1/2St + 1/\sqrt{2St}) + i/\sqrt{2St})$, $\lambda_2 \approx -(1/2St + 1/\sqrt{2St}) - i/\sqrt{2St})$, and $z \approx x + iy - gSt/\Omega^2$.

Let us apply the boundary conditions. The vortex center is located at (0,0). The particles are initially located at $z(0) \approx x_0 + iy_0 - gSt/\Omega^2 = \tilde{x}_0 + iy_0$, $\tilde{x}_0 = x_0 - gSt/\Omega^2$, with the initial velocity of $\dot{z}(0) = 0$. Using these boundary conditions we have $A_1 + A_2 = z(0)$ and $\lambda_1 A_1 + \lambda_2 A_2 = 0$, resulting

$$A_2 = \frac{\lambda_1 z(0)}{\lambda_1 - \lambda_2} \tag{22}$$

and

$$A_1 = -\frac{\lambda_2 z(0)}{\lambda_1 - \lambda_2} \tag{23}$$

Therefore, the general solution for z becomes

$$z(t) = \frac{z(0)}{\lambda_2 - \lambda_1}\left(\lambda_2 e^{\lambda_1 t} - \lambda_1 e^{\lambda_2 t}\right) \tag{24}$$

We separate *z* into real and imaginary parts as

$$x(t) = \mathrm{Re}\{z(t)\}$$

$$= \mathrm{Re}\left\{\frac{z(0)}{\lambda_2 - \lambda_1}\left(\lambda_2 e^{\lambda_1 t} - \lambda_1 e^{\lambda_2 t}\right)\right\} \tag{25}$$

$$y(t) = \mathrm{Im}\{z(t)\}$$

$$= \mathrm{Im}\left\{\frac{z(0)}{\lambda_2 - \lambda_1}\left(\lambda_2 e^{\lambda_1 t} - \lambda_1 e^{\lambda_2 t}\right)\right\} \tag{26}$$

Assume the grains touch the tray surface at time *T* so that

$$Y_0 = \mathrm{Im}\left\{\frac{z(0)}{\lambda_2 - \lambda_1}\left(\lambda_2 e^{\lambda_1 T} - \lambda_1 e^{\lambda_2 T}\right)\right\}$$

or

$$-|Y_0| = \tilde{x}_0 \mathrm{Im}\left\{\frac{\lambda_2 e^{\lambda_1 T} - \lambda_1 e^{\lambda_2 T}}{\lambda_2 - \lambda_1}\right\} + y_0 \mathrm{Re}\left\{\frac{\lambda_2 e^{\lambda_1 T} - \lambda_1 e^{\lambda_2 T}}{\lambda_2 - \lambda_1}\right\} \tag{27}$$

since $Y_0$ is negative. The time $t$ required by the grain to touch the tray surface is not so different from the time for free fall a distance $y_0 + |Y_0|$, i.e. $\sqrt{2(|Y_0| + y_0)/g}$. In real situation we have $y_0 + |Y_0| \approx 0.6$ m so that the time for grain to touch the tray surface is around $t \approx$ 0.35, corresponds to $T \approx \Omega t = 2.2$. Using this value, we can prove $\lambda_1 T \ll 1$ and $\lambda_2 T \ll 1$ so that

$$\frac{\lambda_2 e^{\lambda_1 T} - \lambda_1 e^{\lambda_2 T}}{\lambda_2 - \lambda_1} = \frac{\lambda_2 - \lambda_1 - \frac{1}{2}\lambda_1\lambda_2(\lambda_2 - \lambda_1)T^2}{\lambda_2 - \lambda_1} = 1 - \frac{1}{2}\lambda_1\lambda_2 T^2 \tag{28}$$

Substituting Eq. (27) into (26) one has

$$-|Y_0| = \tilde{x}_0 \,\mathrm{Im}\left\{1 - \frac{1}{2}\lambda_1\lambda_2 T^2\right\} + y_0 \,\mathrm{Re}\left\{1 - \frac{1}{2}\lambda_1\lambda_2 T^2\right\}$$

$$= -\frac{1}{2}\tilde{x}_0 T^2 \,\mathrm{Im}\{\lambda_1\lambda_2\} + y_0 - \frac{1}{2}y_0 T^2 \,\mathrm{Re}\{\lambda_1\lambda_2\}$$

or

$$T = \sqrt{\frac{2(|Y_0| + y_0)}{\tilde{x}_0 \,\mathrm{Im}\{\lambda_1\lambda_2\} + y_0 \,\mathrm{Re}\{\lambda_1\lambda_2\}}} \tag{29}$$

The horizontal position of the grain when touching the tray surface is

$$x(\tau) \approx \tilde{x}_0 \,\mathrm{Re}\left\{1 - \frac{1}{2}\lambda_1\lambda_2 T^2\right\} - y_0 \,\mathrm{Im}\left\{1 - \frac{1}{2}\lambda_1\lambda_2 T^2\right\}$$

$$\approx \tilde{x}_0\left(1 - \frac{T^2}{2}\mathrm{Re}\{\lambda_1\lambda_2\}\right) - \frac{1}{2}y_0 T^2 \,\mathrm{Im}\{\lambda_1\lambda_2\} \tag{30}$$

Let us further perform the following approximation to obtain final result: $\lambda_{1r} \approx -1/2St + 1/\sqrt{2St}$, $\lambda_{1i} \approx 1/\sqrt{2St}$, $\lambda_{2r} \approx -1/2St - 1/\sqrt{2St}$, $\lambda_{2i} \approx -1/\sqrt{2St}$, $\lambda_{1r}\lambda_{2r} - \lambda_{1i}\lambda_{2i} \approx 1/4St^2$, $\lambda_{1r}\lambda_{2i} + \lambda_{1i}\lambda_{2r} = -1/St$ so that

$$\mathrm{Re}\{\lambda_1\lambda_2\} = \lambda_{1r}\lambda_{2r} - \lambda_{1i}\lambda_{2i} \approx \frac{1}{4St^2} \tag{31}$$

$$\text{Im}\{\lambda_1\lambda_2\} = \lambda_{1r}\lambda_{2i} + \lambda_{1i}\lambda_{2r} \approx -\frac{1}{St} \tag{32}$$

Substituting Eqs. (31) and (32) into Eqs. (29) and (30) we have

$$T \approx \sqrt{\frac{2(|Y_0| + y_0)}{-\tilde{x}_0/St + y_0/4St^2}} \tag{33}$$

and

$$\tilde{x}(T) \approx \tilde{x}_0\left(1 - \frac{T^2}{8St^2}\right) - y_0\frac{T^2}{2St} \tag{34}$$

From Eq. (34) we can obtain the relationship between the x-coordinates of the n-th flapping, $\tilde{x}_{0,n}$ and (n+1)-th flapping, $\tilde{x}_{0,n+1}$, as

$$\tilde{x}_{0,n+11} \approx \tilde{x}_{0,n}\left(1 - \frac{T^2}{8St^2}\right) - y_0\frac{T^2}{2St} \tag{35}$$

The difference in the x-positions between the n-th and (n+1)-th flapping becomes

$$\Delta\tilde{x}_{n+1} = \tilde{x}_{0,n+1} - \tilde{x}_{0,n} \approx -\tilde{x}_{0,n}\frac{T^2}{8St^2} - y_0\frac{T^2}{2St} \tag{36}$$

Based on the definition of $\tilde{x}$ we can easily show that $\Delta\tilde{x}_{n+1} = \Delta x_{n+1}$ so that Eq. (36) can be written as

$$\Delta x_{n+1} = -x_{0,n}\frac{T^2}{8St^2} - (y_0 - g/2\Omega^2)\frac{T^2}{2St} \tag{37}$$

The grain may move to the negative or positive direction, depends on $x_0$, and $y_0$. Based on Eq. (37) the grain move to the positive direction if $\Delta x_{n+1} > 0$, resulting

$$y_0 < -\frac{x_{0,n}}{4St} + \frac{g}{2\Omega^2} \tag{38}$$

To the contrary, the grain move to negative of positive direction when

$$y_0 > -\frac{x_{0,n}}{4St} + \frac{g}{2\Omega^2} \tag{39}$$

Based on Eqs. (37) and (38) we can draw a phase diagram showing the regions where the grain move to the positive direction (toward the tray front edge) or to the opposite direction. The diagram is shown in Fig. 9. The two regions are separated by a line with a slope $-1/4St$ and cross the vertical axis at $g/2\Omega^2$. Using the definition in Eq. (14) we obtain the line slope is

$$slope = -\frac{9\mu}{\rho_p \Omega \phi^2} \tag{40}$$

Thus for a certain granular matter, the slope is inversely proportional to square of diameter.

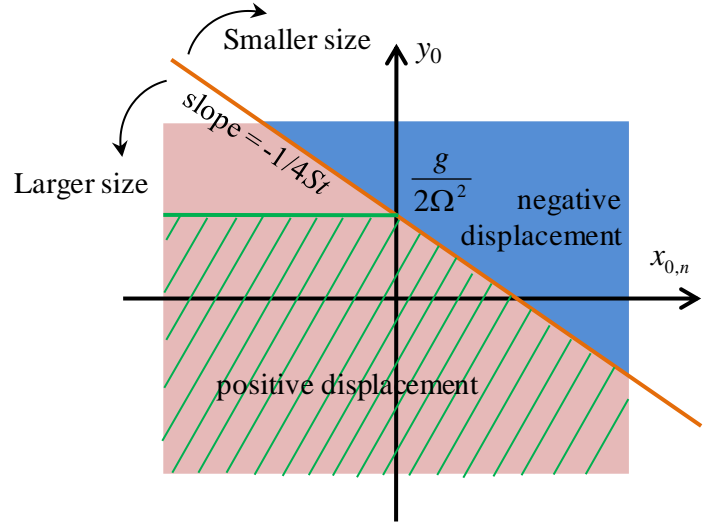

**Fig. 9** Phase diagram showing the regions where the grain move to the positive direction (below the brown line) or to the negative direction (above the brown line) along the horizontal coordinate. The green shaded region is the location of states when the grain can leave the tray by jumping the free end.

Let us further process Eq. (37) as following. This equation can be written as

$$\frac{x_{n+1} - x_n}{(n+1) - n} = -x_{0,n} \frac{T^2}{8St^2} - \left(y_0 - g/2\Omega^2\right)\frac{T^2}{2St}$$

or

$$\frac{\Delta x_n}{\Delta n} = -x_{0,n}\frac{T^2}{8St^2} - (y_0 - g/2\Omega^2)\frac{T^2}{2St}$$

(41)

Equation (41) can be converted into continuous form as

$$\frac{dx_0(n)}{dn} = -\frac{T^2}{8St^2}\left(x_0(n) + 4St(y_0 - g/2\Omega^2)\right)$$

or

$$\frac{dx_0(n)}{\left(x_0(n) + 4St(y_0 - g/2\Omega^2)\right)} = -\frac{T^2}{8St^2}dn \tag{42}$$

Integrating from $n = 0$, $x_0$ to arbitrary $n$ and $x_0(n)$ one obtains

$$x_0(n) = x_0 \exp\left(-\frac{T^2}{8St^2}n\right) - 4St\left(y_0 - \frac{g}{2\Omega^2}\right)\left(1 - \exp\left(-\frac{T^2}{8St^2}n\right)\right) \tag{43}$$

It is clear from Eq. (43) that when $n = 0$ one has $x_0(0) = x_0$, as expected. When $n \to \infty$ one has $x_0(\infty) \to -4St(y_0 - g/2\Omega^2)$. Thus $-4St(y_0 - g/2\Omega^2)$ is the upper bound of $x_0(n)$. This is consistent with the phase diagram in Fig. 9.

As mentioned above, the vortex center is located at around the free end of the tray. Therefore, the entire surface of the tray is located at $x < 0$. To make the grains jump out the tray at the free end, the x coordinate must be positive. It is achieved when $y_0 - g/2\Omega^2 < 0$ or $y_0 < g/2\Omega^2$. The green shaded region in Fig. 9 is the states where the grain can leave the tray by jumping the free end.

As also mention above, the Stokes number in our case is around 348. The time required by the grain to touch the tray surface is less than one second and the number of flapping is usually less than ten times so that $T^2n/8St^2 \ll 1$. Therefore, Eq. (43) can be approximated as

$$x_0(n) \approx x_0 - 4St\left(y_0 - \frac{g}{2\Omega^2}\right)\left(\frac{T^2}{8St^2}n\right)$$

$$\approx x_0 - \frac{T^2}{2St}\left(y_0 - \frac{g}{2\Omega^2}\right)n \tag{44}$$

Equation (44) shows that the horizontal position of the grains increases linearly with the number flapping.

**CONCLUSION**

We have investigated experimentally the process of rice winnowing and develop a model to explain the process of segregation of rice grains having different size or density. We also demonstrated numerically that the effectiveness of segregation is strongly depended on the different in grain sizes (for grain from the same material), the initial position of the grain, and the angular velocity of the vortex generated by flapping the tray. Finally, we also obtained a phase diagram describing different final conditions of winnowing process (either the small grains move towards the tray fee end or move toward the inner end of the tray, able or unable to leave the tray at the free end). The result can be useful to design a new method in separating grains based on size or density.